\documentclass[10pt,aps,twocolumn,superscriptaddress,longbibliography, prd,preprintnumbers]{revtex4-2}
\usepackage{array}
\usepackage{natbib}
\usepackage{amssymb}
\usepackage{amsmath}
\usepackage[utf8]{inputenc}
\usepackage{graphicx}
\usepackage{multirow}
\usepackage{xcolor}
\usepackage{color}
\usepackage[normalem]{ulem}
\usepackage{lipsum,afterpage}
\usepackage{hyperref} 
\usepackage[nameinlink]{cleveref}
\Crefname{appsec}{Appendix}{Appendices}
\usepackage{tensor}
\usepackage{mathtools}
\usepackage{physics}
\usepackage{makecell}
\usepackage{cellspace}

\hypersetup{colorlinks=true, linkcolor=blue, citecolor=blue, urlcolor=blue, linktoc=all, linktocpage=true}

\graphicspath{{./figures/}{./}}

\newcommand{\Chr}[3]{\Gamma{^#1_{#2#3}}}

\newcommand{\idx}[1]{\indices{#1}}
\newcommand{\ud}[2]{\indices{^{#1}_{#2}}}
\newcommand{\du}[2]{\indices{_{#1}^{#2}}}
\newcommand{\dud}[3]{_{#1}{}^{#2}{}_{#3}}

\newcommand{\Real}[0]{\mathbb{R}}
\newcommand{\M}[0]{\mathcal{M}}

\newcommand{\sLC}[3]{\left\{\begin{smallmatrix}#1 \\ #2#3\end{smallmatrix}\right\}}

\definecolor{revcolor}{HTML}{000000}
\newcommand{\revision}[1]{{\color{revcolor}#1}}

\begin{document}
\preprint{YITP-26-100}

\title{Light propagation and intensity transport in metric--affine geometry}

\author{Antonio De Felice}
\email{antonio.defelice@yukawa.kyoto-u.ac.jp}
\affiliation{Center for Gravitational Physics and Quantum Information, Yukawa Institute for Theoretical Physics, Kyoto University, Kyoto 606-8502, Japan}

\author{Lavinia~Heisenberg}
\email{heisenberg@thphys.uni-heidelberg.de}
\affiliation{Institut für Theoretische Physik, Ruprecht-Karls-Universität Heidelberg, Philosophenweg 16, 69120 Heidelberg, Germany}

\author{Gonzalo J. Olmo}
\email{gonzalo.olmo@ific.uv.es}
\affiliation{Instituto de Física Corpuscular (IFIC), CSIC-Universitat de València, Spain}
\affiliation{Departamento de Física, Universidade Federal do Ceará, Campus do Pici, 60455-760 Fortaleza, Ceará, Brazil}

\author{Carlos Pastor-Marcos}
\email{pastor\_c@thphys.uni-heidelberg.de}
\affiliation{Institut für Theoretische Physik, Ruprecht-Karls-Universität Heidelberg, Philosophenweg 16, 69120 Heidelberg, Germany}

\date{\today}

\begin{abstract}
We study electromagnetic wave propagation in metric--affine geometries, where torsion and non-metricity may be present and the coupling between electromagnetism and spacetime is no longer unique. Rather than choosing a particular coupling prescription a priori, we construct electromagnetic sectors that preserve standard $U(1)$ gauge invariance and projective invariance of the affine connection as guiding symmetry principles. We introduce two representative models, one in which the Maxwell term is dressed by a scalar prefactor built from non-Riemannian invariants, and another in which the kinetic term is modified by a rank--four constitutive tensor acting as an anisotropic medium. We derive their geometric--optics limits and show that their couplings can modify the effective light cone, change the relation between field amplitude and intensity, induce polarization-dependent propagation, and generate birefringence and mode mixing. These results thus provide the formal basis for a broader phenomenological study connecting torsion and non-metricity with electromagnetic observables in concrete metric--affine backgrounds, including black-hole imaging, birefringent lensing, polarization observables, and departures from photon-number conservation.
\end{abstract}

\maketitle


\section{Introduction} 
\label{sec:Introduction}
General Relativity (GR) provides a remarkably successful description of gravitational phenomena across a vast range of scales. In its standard formulation, spacetime geometry is entirely determined by the metric, and the affine connection is uniquely fixed to be the Levi--Civita one. This structure implies the absence of torsion and non-metricity and, consequently, a single natural notion of parallel transport and inertial motion. The trajectory of a free test particle can then be characterized in two equivalent ways: on the one hand, as a geodesic arising from the extremization of the spacetime interval, $S=\int ds$; on the other hand, as an autoparallel, defined by requiring that the tangent vector be parallel transported along the curve, $\nabla_{\dot x}\dot x=0$.
For the Levi--Civita connection these two definitions coincide: geodesics and autoparallels are described by the same equation, and free-falling particles follow metric geodesics. However, this equivalence is a special feature of Riemannian geometry, and in more general geometrical frameworks the situation changes qualitatively \cite{Heisenberg:2023lru,Heisenberg:2026ess,Delhom:2020vpe,Avalos:2016unj}.

In theories such as metric--affine gravity, teleparallel gravity, or symmetric teleparallel gravity, the affine connection is not determined by the metric alone. Instead, curvature, torsion, and non-metricity may appear as independent geometrical quantities \cite{Heisenberg:2018vsk, BeltranJimenez:2019esp,Heisenberg:2023lru}. In these settings, the distinction between geodesics and autoparallels becomes physically meaningful, since the former are still determined solely by extremizing $ds$, whereas the latter now depend on the full affine connection through the condition of parallel transport \cite{Heisenberg:2026ess,Olmo:2022ops,Heisenberg:2023lru}. The relation between particle dynamics and spacetime geometry is therefore less straightforward than in GR, and the notion of ``natural'' or inertial motion is no longer unique. Particle motion in metric--affine spacetimes, including the effects of torsion and non-metricity, has been studied from complementary perspectives \cite{Puetzfeld:2014qba,Iosifidis:2023eom}. More recently, \cite{Heisenberg:2026ess} addressed the specific question of whether autoparallels can arise from a reparametrization-invariant scalar worldline action, showing this for torsion-free affine connections with arbitrary non-metricity, while the extension of this construction to torsional and mixed cases remains open. This illustrates a broader ambiguity that arises in non-Riemannian geometries: whether physical trajectories should be tied to the metric structure, to the affine structure, or to the specific field equations of the matter sector under consideration. Different theoretical frameworks adopt different viewpoints, and in general the choice may affect physical predictions. 

A closely related problem arises when considering how matter fields propagate in such geometries. In GR coupled to Maxwell electrodynamics, electromagnetic waves in the geometric--optics (GO) limit travel along null geodesics of the spacetime metric, while in nonlinear theories of electrodynamics they follow geodesics of a different effective metric \cite{Dolan:2017zgu,Novello:1999pg,Bunney:2019yzx}. Beyond GR, once the connection is allowed to deviate from the Levi--Civita form, the coupling of electromagnetism to the geometry also becomes ambiguous \cite{Hehl:1999bt,BeltranJimenez:2020sih,Delhom:2020hkb,Nieh:2017zxr}. As a result, the relation between photon trajectories, the spacetime metric, and the affine connection requires careful reconsideration. The field equations of the resulting theory will thus tell us what the preferred paths of light rays are, without any need to impose or guess their form.

In this work we investigate the propagation of electromagnetic fields in metric--affine geometries from this perspective. Rather than imposing a specific coupling between electromagnetism and the connection from the outset, which is an unnecessary form of violence, we analyze the problem systematically by identifying the ambiguities that arise when the connection is treated as an independent geometric object. 
We then use standard $U(1)$ gauge invariance and projective invariance as symmetry principles to guide the construction of consistent electromagnetic sectors and study the resulting photon propagation in the GO limit. We show that the non-Riemannian structure of spacetime can modify the light cone, change the intensity along a beam, induce birefringence, or generate polarization-dependent propagation, and discuss how such modifications may lead to observable effects \cite{dePaula:2025emo,Bunney:2019yzx,Wang:2024dkn}. 

The present work constitutes the first in a series of papers devoted to connecting these theories with observations. Here we focus on the more formal aspects, identifying consistent couplings and deriving the corresponding propagation equations, whereas subsequent works will investigate concrete phenomenological signatures, including possible modifications of black-hole shadows, birefringent lensing, polarization-dependent propagation, and \revision{apparent} photon-number non-conservation induced by an effective exchange between electromagnetic radiation and the underlying non-Riemannian geometry.

\section{Electromagnetism in metric--affine geometry}
\label{sec:EM_MAG}

\subsection{Metric--affine geometry and conventions}
\label{subsec:MAG_setup}

We work on a differentiable spacetime manifold $\M$ equipped with a metric $g_{\mu\nu}$ and an independent affine connection $\Chr{\alpha}{\mu}{\nu}$. The curvature tensor associated with the connection is defined as
\begin{equation}
    R\ud{\alpha}{\beta\mu\nu} \coloneqq \partial_{\mu}\Chr{\alpha}{\nu}{\beta}
    - \partial_{\nu}\Chr{\alpha}{\mu}{\beta} 
    + \Chr{\alpha}{\mu}{\lambda}\Chr{\lambda}{\nu}{\beta}
    - \Chr{\alpha}{\nu}{\lambda}\Chr{\lambda}{\mu}{\beta},
\label{eq:RiemannTensor}
\end{equation}
and the Ricci tensor and scalar are the contractions
\begin{equation}
    R_{\alpha\beta} \coloneqq R\idx{^\mu_{\alpha\mu\beta}},
    \qquad
    R \coloneqq R\idx{^\alpha_\alpha}.
\label{eq:RicciDefinitions}
\end{equation}
In contrast to GR, the connection is not assumed to be symmetric nor metric compatible. The antisymmetric part of the connection defines the torsion tensor
\begin{equation}
    T\ud{\alpha}{\mu\nu} \coloneqq 2\,\Chr{\alpha}{[\mu}{\nu]},
\label{eq:TorsionTensor}
\end{equation}
while the failure of the connection to preserve the metric defines the non-metricity tensor
\begin{equation}
    Q\idx{_{\alpha\mu\nu}} \coloneqq \nabla_{\alpha} g_{\mu\nu}.
\label{eq:NonMetricityTensor}
\end{equation}
From this definition it follows that raising indices introduces a minus sign, $\nabla_{\alpha}g^{\mu\nu}=-Q\du{\alpha}{\mu\nu}$.

Equivalently, the torsion and curvature tensors can be obtained from the commutator of covariant derivatives acting on a scalar $\phi$ and a vector $A^{\mu}$, respectively:
\begin{align}
    [\nabla_{\mu},\nabla_{\nu}]\phi &= -\,T\ud{\alpha}{\mu\nu}\,\partial_{\alpha}\phi, \nonumber\\[1ex]
    [\nabla_{\mu},\nabla_{\nu}]A^{\rho} &= R\idx{^\rho_{\alpha\mu\nu}}A^{\alpha} - T\ud{\alpha}{\mu\nu}\nabla_{\alpha}A^{\rho},
\label{eq:CommutatorsCovDeriv}
\end{align}
and their independent traces are
\begin{equation}
    T_\alpha\coloneqq T\ud{\mu}{\alpha\mu},\qquad
    Q_\alpha\coloneqq Q\du{\alpha\mu}{\mu},\qquad
    \tilde Q_\alpha\coloneqq Q\du{\mu\alpha}{\mu}.
\label{eq:TandQTraces}
\end{equation}

It is often convenient to decompose the affine connection into the Levi--Civita part of the metric $\sLC{\alpha}{\mu}{\nu}$ and a distortion tensor encoding the contributions from torsion and non-metricity,
\begin{equation}
    \Chr{\alpha}{\mu}{\nu} = \sLC{\alpha}{\mu}{\nu} + \Omega\ud{\alpha}{\mu\nu},
\label{eq:ConnectionDecomposition}
\end{equation}
with
\begin{align}
    \sLC{\alpha}{\mu}{\nu} &= \frac{1}{2}\, g^{\alpha\lambda} \left( \partial_{\mu} g_{\nu\lambda} + \partial_{\nu} g_{\mu\lambda} - \partial_{\lambda} g_{\mu\nu} \right),\nonumber\\[1ex]
    \Omega\ud{\alpha}{\mu\nu} &= K\ud{\alpha}{\mu\nu}(T) + L\ud{\alpha}{\mu\nu}(Q).
\label{eq:DistortionDefinition}
\end{align}
In particular, $K\ud{\alpha}{\mu\nu}$ and $L\ud{\alpha}{\mu\nu}$ are the so--called contorsion and disformation tensors, defined as
\begin{align}
    K\ud{\alpha}{\mu\nu} &\coloneqq \frac{1}{2}T\ud{\alpha}{\mu\nu} + T\dud{(\mu}{\alpha}{\nu)},\nonumber\\[1ex]
    L\ud{\alpha}{\mu\nu} &\coloneqq \frac{1}{2}Q\ud{\alpha}{\mu\nu} - Q\dud{(\mu}{\alpha}{\nu)}.
\label{eq:ContorsionDisformation}
\end{align}
They carry the contributions of torsion and non-metricity, respectively. $K_{\alpha\mu\nu}$ is antisymmetric under the exchange of the first and third indices, while $L\ud{\alpha}{\mu\nu}$ is symmetric in the last pair. The distortion tensor $\Omega\ud{\alpha}{\mu\nu}$ has no particular symmetry properties.

Useful relations are
\begin{align}
    Q_{\alpha\mu\nu}&=-L_{\mu\alpha\nu}-L_{\nu\alpha\mu},\nonumber\\[1ex]
    T_{\alpha\mu\nu}&=K_{\alpha\mu\nu}-K_{\alpha\nu\mu}.
\label{eq:TQ_from_KL}
\end{align}

In the following we denote by $\nabla_\mu$ the covariant derivative associated with the full affine connection, while $\mathring{\nabla}_\mu$ refers to the Levi--Civita derivative. Curvature quantities computed with the Levi--Civita connection are denoted by calligraphic symbols, namely $\mathcal{R}\idx{^\rho_{\alpha\mu\nu}}$, $\mathcal{R}_{\alpha\beta}$ and $\mathcal{R}$.

\subsection{Maxwell theory and GO in and beyond GR}
\label{subsec:Maxwell_GO}

Before considering the general metric--affine case, it is useful to recall the standard formulation of electromagnetism in GR. The electromagnetic field strength tensor is defined as
\begin{equation}
    F_{\mu\nu} \coloneqq \partial_\mu A_\nu - \partial_\nu A_\mu ,
\label{eq:FaradayTensor}
\end{equation}
which is invariant under $U(1)$ gauge transformations.
In particular, the electromagnetic potential can be interpreted as a one-form $A=A_\mu dx^\mu$, so that the Faraday tensor arises naturally as the exterior derivative $F:=dA$. This construction does not require any affine connection and is therefore well-defined independently of the underlying spacetime geometry \cite{Hehl:1999bt}. In Riemannian geometry, this is equivalent to writing 
\begin{equation}
    \mathring{F}_{\mu\nu} \coloneqq \mathring{\nabla}_\mu A_\nu - \mathring{\nabla}_\nu A_\mu ,
\label{eq:FLC}
\end{equation}
since the Levi--Civita connection is torsionless and the standard $U(1)$ gauge symmetry is thus preserved. Writing the field strength in this form is convenient in curved spacetime, since it allows one to exploit standard tensorial identities involving covariant derivatives and makes the comparison with more general affine connections conceptually more transparent \cite{Dolan:2017zgu,Maggiore:2007ulw}. For this reason, we adopt the representation \labelcref{eq:FLC} in what follows.

The dynamics of the electromagnetic field is governed by the Maxwell action
\begin{equation}
    S_{\text{EM}} = -\frac{1}{4}\int d^4x \,\sqrt{-g}\,\mathring{F}_{\mu\nu}\mathring{F}^{\mu\nu},
\label{eq:MaxwellAction}
\end{equation}
which yields the vacuum Maxwell equations
\begin{equation}
    \mathring{\nabla}_\mu \mathring{F}^{\mu\nu} = 0 .
\label{eq:MaxwellEquations}
\end{equation}

To study the propagation of matter fields in the high--frequency regime we employ the GO approximation \cite{Maggiore:2007ulw,Misner:1973prb,Poisson_Will_2014,Dolan:2017zgu}. In this limit the field is decomposed into a rapidly oscillating exponential term and a slowly varying amplitude. We consider the Ansatz
\begin{equation}
    A_\mu = a_\mu\,e^{i\omega\Psi},
\label{eq:GO_Ansatz}
\end{equation}
where $a_\mu$ is a slowly varying amplitude, $\omega\gg1$ is a large parameter controlling the GO expansion, and $\Psi$ is a scalar phase function.

The wave covector is defined as the gradient of the phase,
\begin{equation}
    k_\mu \coloneqq \mathring{\nabla}_\mu\Psi,
\label{eq:WaveCovector}
\end{equation}
and we introduce the divergence
\begin{equation}
    \mathring{\nabla}_\mu k^\mu = \theta ,
\label{eq:ThetaDefinition}
\end{equation}
where $\theta$ denotes the slowly varying expansion of the wave congruence.

In the GO limit derivatives acting on the exponential produce factors of $\omega$, while derivatives of the amplitudes are assumed to be of order $\mathcal{O}(\omega^0)$. Accordingly, the field equations can be expanded in powers of $\omega$. At leading order we retain terms of order $\mathcal{O}(\omega^2)$ and $\mathcal{O}(\omega)$, while quantities such as the background curvature typically scale as $\mathcal{O}(\omega^0)$. In particular, in GR the Ricci tensor satisfies $\mathcal{R}_{\mu\nu}\sim\mathcal{O}(\omega^0)$ and therefore does not contribute at leading order in the GO expansion. At $\mathcal{O}(\omega^2)$ one has:
\begin{equation}
    g^{\mu\nu}k_\mu k_\nu = 0 ,
\label{eq:NullDispersion}
\end{equation}
showing that electromagnetic waves propagate along null geodesics of the spacetime metric.

For later comparison, it is useful to distinguish two natural ways of extending Maxwell theory beyond GR. The first one keeps the standard Maxwell tensor in \labelcref{eq:FLC} and introduces the independent affine structure only through the equations of motion (EoMs), generalizing \labelcref{eq:MaxwellEquations} as
\begin{equation}
    \nabla^\nu \mathring{F}_{\mu\nu} = 0 .
\label{eq:MaxwellLC_genDiv}
\end{equation}
With this prescription, the non-Riemannian corrections arise only through the outer covariant derivative, while the antisymmetric field strength is left unchanged.

A second possibility is to promote the affine structure also in the definition of the field strength and consider instead
\begin{equation}
    \hat F_{\mu\nu} \coloneqq \nabla_\mu A_\nu - \nabla_\nu A_\mu .
\label{eq:FaradayCovariant}
\end{equation}
Expanding the covariant derivatives, one finds
\begin{equation}
    \hat F_{\mu\nu} = \partial_\mu A_\nu - \partial_\nu A_\mu - T\ud{\lambda}{\mu\nu}A_\lambda ,
\label{eq:FaradayWithTorsion}
\end{equation}
so that the field strength itself acquires an explicit dependence on the affine structure. In particular, torsion enters already at the level of $\hat F_{\mu\nu}$ and the standard $U(1)$ gauge symmetry is generically broken \cite{Nieh:2017zxr,Delhom:2020hkb}. Once \labelcref{eq:FaradayCovariant} is inserted into the EoMs, torsion and non-metricity contribute not only through the outer derivative but also through the structure of the field strength itself, leading to substantially different corrections from those obtained from \labelcref{eq:MaxwellLC_genDiv}.

This shows that, once the connection is treated as an independent object, the coupling between electromagnetism and geometry is no longer unique, and different prescriptions for constructing the field strength tensor or defining the covariant derivatives can lead to inequivalent EoMs \cite{Nieh:2017zxr,BeltranJimenez:2020sih,Delhom:2020hkb}. Formulating consistent electromagnetic sectors in metric--affine geometry therefore requires additional guiding principles, which will be discussed in the following sections.

\section{Symmetry-guided construction of electromagnetic models}
\label{sec:symmetry_principles}

Rather than choosing a particular prescription \emph{a priori}, we adopt a symmetry-based approach to constrain the large class of electromagnetic extensions allowed in the presence of torsion and non-metricity. In particular, the construction of viable theories will be guided by
\begin{enumerate}
\item preservation of the standard $U(1)$ gauge symmetry of electromagnetism,
\item invariance under projective transformations of the affine connection.
\end{enumerate}
Since gauge invariance is experimentally well established, we regard it as a fundamental requirement for any viable extension of Maxwell theory \cite{Nieh:2017zxr}. Projective symmetry, on the other hand, is a natural redundancy of the affine connection in metric--affine gravity, and provides a criterion for identifying physically meaningful couplings between matter fields and the non-Riemannian geometry \cite{Janssen:2018exh,Iosifidis:2024ksa}, as we now discuss.

\subsection{Projective symmetry}

A projective transformation of the affine connection is defined as
\begin{equation}
    \Chr{\alpha}{\mu}{\nu} \;\longrightarrow\; \Chr{\alpha}{\mu}{\nu} + \delta^\alpha_\nu \,\xi_\mu ,
\label{eq:projective_transformation}
\end{equation}
where $\xi_\mu$ is an arbitrary one-form. This transformation preserves the unparametrized autoparallel curves associated with the connection \cite{Janssen:2018exh,Sauro:2022hoh,Olmo:2022ops,Bejarano:2019zco}. Indeed, consider the autoparallel equation
\begin{equation}
    \ddot x^\alpha + \Chr{\alpha}{\mu}{\nu}\,\dot x^\mu \dot x^\nu = 0 ,
\label{eq:autoparallel_equation}
\end{equation}
where $\dot x^\mu\equiv\dv*{x^\mu}{\lambda}$ and $\lambda$ denotes an affine parameter along the curve. Under the transformation \labelcref{eq:projective_transformation}, the left-hand side becomes
\begin{equation}
    \ddot x^\alpha+\Chr{\alpha}{\mu}{\nu}\,\dot x^\mu \dot x^\nu 
    \quad\longrightarrow\quad
    \ddot x^\alpha+\Chr{\alpha}{\mu}{\nu}\,\dot x^\mu \dot x^\nu+(\xi_\mu \dot x^\mu)\,\dot x^\alpha .
\label{eq:autoparallel_projective}
\end{equation}
The additional term is proportional to the tangent vector $\dot x^\alpha$ and can therefore be absorbed by a redefinition of the parameter $\lambda$. As a consequence, the geometric trajectory remains unchanged and only its parametrization is modified. In this sense, projective transformations relate different affine connections that describe the same unparametrized autoparallel curves, and therefore represent a gauge redundancy of the affine structure.

The torsion and non-metricity tensors defined in \labelcref{eq:TorsionTensor} and \labelcref{eq:NonMetricityTensor} transform under \labelcref{eq:projective_transformation} as
\begin{subequations}
\label{eq:ProjShiftTQ}
\begin{align}
    T\ud{\alpha}{\mu\nu} &\longrightarrow T\ud{\alpha}{\mu\nu} + \delta^\alpha_\nu \,\xi_\mu - \delta^\alpha_\mu \,\xi_\nu , \\[1ex]
    Q_{\alpha\mu\nu} &\longrightarrow Q_{\alpha\mu\nu} - 2\,\xi_\alpha g_{\mu\nu}.
\end{align}
\end{subequations}
Equivalently, using the contorsion and disformation tensors introduced in \labelcref{eq:ContorsionDisformation}, their transformations read
\begin{subequations}
\label{eq:ProjShiftKL}
\begin{align}
    K\ud{\alpha}{\mu\nu} &\longrightarrow K\ud{\alpha}{\mu\nu} + g_{\mu\nu}\,\xi^\alpha - \delta^\alpha_\mu\,\xi_\nu , \\[1ex]
    L\ud{\alpha}{\mu\nu} &\longrightarrow L\ud{\alpha}{\mu\nu} - g_{\mu\nu}\,\xi^\alpha + \delta^\alpha_\mu\,\xi_\nu + \delta^\alpha_\nu\,\xi_\mu .
\end{align}
\end{subequations}

Consequently, the traces defined in \labelcref{eq:TandQTraces} transform in $n\equiv\dim\mathcal{M}$ dimensions according to
\begin{subequations}
\label{eq:ProjTraceShifts}
\begin{align}
    T_\mu &\longrightarrow T_\mu +(n-1)\,\xi_\mu , \\[1ex]
    Q_\mu &\longrightarrow Q_\mu - 2n\,\xi_\mu , \\[1ex]
    \tilde Q_\mu &\longrightarrow \tilde Q_\mu -2\,\xi_\mu .
\end{align}
\end{subequations}

These relations show that the trace parts of torsion and non-metricity carry the projective mode and can therefore be shifted arbitrarily by $\xi_\mu$. It is convenient to make this structure explicit by decomposing both tensors into traceful parts carrying the projective mode, $T_{\mathrm t}$ and $Q_{\mathrm t}$, and tracefree parts, $T_{\mathrm f}$ and $Q_{\mathrm f}$, which are projectively invariant.

For the torsion tensor one writes
\begin{subequations}
\begin{align}
    T\ud{\alpha}{\mu\nu} &= T_{\mathrm t}\,\ud{\alpha}{\mu\nu} + T_{\mathrm f}\,\ud{\alpha}{\mu\nu}, \\[1ex]
    T_{\mathrm t}\,\ud{\alpha}{\mu\nu} &\coloneqq -\frac{1}{n-1}\left(\delta^\alpha_\mu\,T_\nu-\delta^\alpha_\nu\,T_\mu\right), \\[1ex]
    T_{\mathrm f}\,\ud{\alpha}{\mu\nu} &\coloneqq T\ud{\alpha}{\mu\nu} - T_{\mathrm t}\,\ud{\alpha}{\mu\nu}.
\end{align}
\label{eq:TtfDecomposition}
\end{subequations}
Using \labelcref{eq:ProjShiftTQ} and \labelcref{eq:ProjTraceShifts}, one finds indeed
\begin{equation}
    \delta T_{\mathrm t}\,\ud{\alpha}{\mu\nu} = \delta^\alpha_\nu\,\xi_\mu-\delta^\alpha_\mu\,\xi_\nu,
    \qquad \delta T_{\mathrm f}\,\ud{\alpha}{\mu\nu}=0 .
\label{eq:ProjShiftTf}
\end{equation}

For the non-metricity tensor the analogous decomposition reads
\begin{subequations}
\begin{align}
    Q_{\alpha\mu\nu} &= Q_{\mathrm{t},\,\alpha\mu\nu} + Q_{\mathrm{f},\,\alpha\mu\nu}, \\[1ex]
    Q_{\mathrm{t},\,\alpha\mu\nu} &\coloneqq \frac{1}{n}\,g_{\mu\nu}\,Q_\alpha, \\[1ex]
    Q_{\mathrm{f},\,\alpha\mu\nu} &\coloneqq Q_{\alpha\mu\nu} - Q_{\mathrm{t},\,\alpha\mu\nu}.
\end{align}
\label{eq:QtfDecomposition}
\end{subequations}
This implies
\begin{equation}
    \delta Q_{\mathrm{t},\,\alpha\mu\nu}=-2\,\xi_\alpha g_{\mu\nu},
    \qquad \delta Q_{\mathrm{f},\,\alpha\mu\nu}=0 .
\label{eq:ProjShiftQf}
\end{equation}
Furthermore, $T_{\mathrm f}$ and $Q_{\mathrm f}$ satisfy
\begin{subequations}
\label{eq:TfQfTraces}
\begin{align}
    T_{\mathrm{f},\,\alpha}\coloneqq\, & T_{\mathrm f}\,\ud{\mu}{\alpha\mu} = 0 , \\[1ex]
    Q_{\mathrm{f},\,\alpha}\coloneqq\, & Q_{\mathrm f}\,\du{\alpha\mu}{\mu} = 0 , \\[1ex]
    \tilde Q_{\mathrm{f},\,\alpha}\coloneqq\, & Q_{\mathrm f}\,\du{\mu\alpha}{\mu}=\tilde Q_\alpha-\frac{1}{n}Q_\alpha .
\end{align}
\end{subequations}
Thus, while $T_{\mathrm f}$ is fully traceless, $Q_{\mathrm f}$ still contains the projectively invariant trace combination $\tilde Q_{\mathrm f,\alpha}$.

Matter couplings depending explicitly on the projective mode would therefore introduce an unphysical sensitivity to this gauge redundancy \cite{Olmo:2013lta,Janssen:2018exh,Bejarano:2019zco,Olmo:2022ops}. For this reason, we require the electromagnetic sector to be built only from projectively invariant combinations of the geometric tensors. In practice, this means that the couplings can depend on $T_{\mathrm f}$, $Q_{\mathrm f}$, and scalar invariants constructed from them. In what follows we also specialize to four spacetime dimensions, $n=4$.

\subsection{Projectively and $U(1)$-invariant electromagnetic models}
\label{subsec:PI_U1_models}

We introduce two representative classes of electromagnetic models satisfying the two symmetry principles above. These models are designed to capture the main qualitative effects that a non-Riemannian background can induce on light propagation, and which may be relevant for future phenomenological studies. In both cases, the Faraday tensor is taken to be the Levi--Civita one, $\mathring{F}_{\mu\nu}$, so that $U(1)$ gauge invariance remains manifest.

One natural possibility is to modify the Maxwell term by a scalar prefactor $\Xi$ built from scalar invariants of the projectively invariant tensors introduced above. In this case the electromagnetic Lagrangian takes the form
\begin{equation}
    \mathcal{L}_{\Xi} \coloneqq -\frac{1}{4}\,\Xi\,\mathring{F}_{\mu\nu}\mathring{F}^{\mu\nu}.
\label{eq:LXi_general}
\end{equation}
Restricting to parity-even, derivative-free terms up to quadratic order in
$T_{\mathrm f}$ and $Q_{\mathrm f}$, the most general Ansatz reads
\begin{align}
    \Xi &= 1
    + c_{T_1}\,T_{\mathrm f}^{\;\alpha\mu\nu}T_{\mathrm f,\,\alpha\mu\nu}
    + c_{T_2}\,T_{\mathrm f}^{\;\alpha\mu\nu}T_{\mathrm f,\,\mu\alpha\nu} \nonumber\\[1ex]
    &\quad
    + c_{Q_1}\,Q_{\mathrm f}^{\;\alpha\mu\nu}Q_{\mathrm f,\,\alpha\mu\nu} 
    + c_{Q_2}\,Q_{\mathrm f}^{\;\alpha\mu\nu}Q_{\mathrm f,\,\mu\alpha\nu} \nonumber\\[1ex]
    &\quad 
    + c_{Q_3}\,\tilde Q_{\mathrm f}^{\,\alpha}\tilde Q_{\mathrm f,\,\alpha}
    + c_{TQ}\,T_{\mathrm f}^{\;\alpha\mu\nu}Q_{\mathrm f,\,\mu\alpha\nu},
\label{eq:Xi_general}
\end{align}
with $\{c_{T_1},c_{T_2},c_{Q_1},c_{Q_2},c_{Q_3},c_{TQ}\}\in\Real$. The constant term reproduces standard Maxwell electrodynamics, while the remaining contributions encode the coupling of the electromagnetic sector to the projectively invariant parts of torsion and non-metricity. 

Another natural possibility is to modify the Maxwell kinetic term through a rank--four constitutive tensor $\chi^{\mu\nu\alpha\beta}$, leading to the Lagrangian
\begin{equation}
    \mathcal{L}_{\chi} \coloneqq -\frac{1}{4}\,\chi^{\mu\nu\alpha\beta}\,\mathring{F}_{\mu\nu}\mathring{F}_{\alpha\beta}.
\label{eq:Lchi_general}
\end{equation}
Since $\mathring{F}_{\mu\nu}$ is antisymmetric, the constitutive tensor can be taken to satisfy the usual symmetries
\begin{equation}
    \chi^{\mu\nu\alpha\beta}
    =
    -\,\chi^{\nu\mu\alpha\beta}
    =
    -\,\chi^{\mu\nu\beta\alpha}
    =
    \chi^{\alpha\beta\mu\nu}.
\label{eq:chi_symmetries}
\end{equation}
A minimal projectively invariant Ansatz up to quadratic order in $T_{\mathrm f}$ and $Q_{\mathrm f}$ can then be written in a form that manifestly respects these symmetries as
\begin{align}
    \chi^{\mu\nu\alpha\beta} &= g^{\mu[\alpha}g^{\beta]\nu} 
    + \bar{c}_T\,T_{\mathrm f}^{\;\delta\mu\nu}T_{\mathrm f\,\delta}{}^{\alpha\beta} \nonumber\\[1ex] 
    &\quad + \frac{\bar{c}_Q}{2} \Big( Q_{\mathrm f\,\delta}{}^{\mu\alpha}Q_{\mathrm f\,}{}^{\delta\nu\beta} - Q_{\mathrm f\,\delta}{}^{\nu\alpha}Q_{\mathrm f\,}{}^{\delta\mu\beta} \Big),
\label{eq:chi_general_symmetric}
\end{align}
with constant coefficients $\{\bar{c}_T,\bar{c}_Q\}\in\Real$. However, the action only probes the projected component $\chi^{[\mu\nu][\alpha\beta]}$ through its contraction with $\mathring{F}_{\mu\nu}\mathring{F}_{\alpha\beta}$, so one may work equivalently with the simpler representative
\begin{equation}
    \chi^{\mu\nu\alpha\beta} = g^{\mu\alpha}g^{\nu\beta} + 
    \bar{c}_T\,T\du{\mathrm{f},\,\delta}{\mu\nu} T_{\mathrm{f}}^{\;\delta\alpha\beta} + 
    \bar{c}_Q\,Q\du{\mathrm{f},\,\delta}{\mu\alpha} Q_{\mathrm{f}}^{\;\delta\nu\beta},
\label{eq:chi_general}
\end{equation}
which reproduces the same projected component and therefore yields the same dynamics, so we adopt \labelcref{eq:chi_general} in what follows. Maxwell theory follows from its first term, while the remaining ones encode the response of the electromagnetic field to the projectively invariant components of the non-Riemannian geometry.

We note that, in contrast to the scalar prefactor model \labelcref{eq:LXi_general}, for which all independent quadratic contractions have been included, the constitutive framework \labelcref{eq:Lchi_general} admits a substantially larger set of tensorial structures. For the purposes of this work, however, the minimal Ansatz \labelcref{eq:chi_general} is already rich enough to exhibit the qualitative features we aim to study, namely a modified dispersion relation, non-trivial intensity evolution, and birefringence, as will be shown in the following sections. Related non-minimal electromagnetic couplings quadratic in torsion have previously been shown to induce an effective axion sector and vacuum birefringence \cite{Itin:2003hr}. Nevertheless, a systematic analysis including all admissible terms is certainly of interest and will be addressed elsewhere.

From the perspective of electrodynamics in media, model \labelcref{eq:Lchi_general} can be interpreted as an effective constitutive medium induced by the non-Riemannian geometry \cite{Hehl:1999bt,Itin:2003hr,Novello:1999pg,Queiruga:2019ike,Lobo:2014nwa}, and in general it is expected to lead to polarization-dependent propagation. The scalar prefactor model \labelcref{eq:LXi_general}, on the contrary, rescales the Maxwell sector without introducing a non-trivial tensorial structure in the kinetic term. Its effect on photon propagation is therefore qualitatively simpler and, in particular, it is not expected to induce polarization mixing at leading order, as will be analyzed in the GO limit below. 
\revision{Formally, \labelcref{eq:LXi_general} is the particular case $\chi^{\mu\nu\alpha\beta}=\Xi\,g^{\mu[\alpha}g^{\beta]\nu}$ of \labelcref{eq:Lchi_general}, and replacing the first term of \labelcref{eq:chi_general_symmetric} with $\Xi\,g^{\mu[\alpha}g^{\beta]\nu}$ gives a constitutive tensor that combines both couplings,\revision{
\begin{equation*}
    \color{revcolor}\chi^{\mu\nu\alpha\beta}=\Xi\,g^{\mu[\alpha}g^{\beta]\nu}+\delta\chi^{\mu\nu\alpha\beta} ,
\end{equation*}
where $\delta\chi^{\mu\nu\alpha\beta}$ denotes the $\bar c_T$ and $\bar c_Q$ terms of \labelcref{eq:chi_general_symmetric}. The GO analysis of \Cref{subsec:GO_constitutive_model} applies directly to this tensor.} We nevertheless study the two models separately, \revision{i.e., the limiting cases $\bar c_T=\bar c_Q=0$ and $\Xi=1$,} because, as we shall see, they isolate different effects: in the scalar model $\Xi$ leaves the light cone unchanged and enters only the amplitude transport, whereas $\bar c_T$ and $\bar c_Q$ deform the light cone and induce birefringence.}
In general, the two theories will serve as our preferred representative models in what follows and provide two distinct realizations of how projectively invariant metric--affine geometry can couple to electromagnetism while preserving the standard $U(1)$ gauge symmetry.

\section{GO analysis of the electromagnetic models}
\label{sec:GO_models}

We now analyze the propagation of electromagnetic waves in the two representative models introduced above. Our goal is to derive their EoMs and study the high--frequency propagation in the GO limit. In particular, we will identify the leading-order dispersion relation, which determines the propagation of the wavefronts, and the next-to-leading transport equations, which govern the evolution of the amplitude, the intensity profile, and, when applicable, the polarization state.

Rather than expanding immediately in terms of the explicit torsion and non-metricity components, it is convenient to keep the equations expressed in terms of the scalar prefactor $\Xi$ and the constitutive tensor $\chi^{\mu\nu\alpha\beta}$. This makes the structure of the propagation equations transparent and avoids unnecessarily lengthy expressions at this stage of the analysis.

\subsection{Scalar-coupled model}
\label{subsec:GO_scalar_model}

For the scalar model \labelcref{eq:LXi_general}, variation with respect to $A_\mu$ yields
\begin{equation}    \Xi\,\mathring{\nabla}_\alpha\mathring{F}\ud{\alpha}{\mu} +(\mathring{\nabla}_\alpha\Xi)\,\mathring{F}\ud{\alpha}{\mu}=0.
\label{eq:EOM_Xi_general}
\end{equation}
We now implement the GO Ansatz \labelcref{eq:GO_Ansatz} and expand the EoMs in powers of the large parameter $\omega$. 

\subsubsection*{GO limit: $\mathcal{O}(\omega^2)$}

Using the standard rules discussed in \Cref{subsec:Maxwell_GO} and imposing the Lorenz gauge $\mathring{\nabla}^\mu A_\mu=0$, the leading contribution arises at order $\mathcal{O}(\omega^2)$.
From \labelcref{eq:EOM_Xi_general} one obtains
\begin{equation}
    \Xi\,k_\alpha k^\alpha\,a_\mu=0 .
\label{eq:GO_Xi_omega2}
\end{equation}
For a generic background $\Xi\neq0$, the dispersion relation is therefore still given by \labelcref{eq:NullDispersion} and the light cone is not deformed at leading GO order. It is convenient to rewrite this relation as
\begin{equation}
    k_{\rm eff}^2 \equiv g_{\rm eff}^{\mu\nu}k_\mu k_\nu=0,
\label{eq:keff_scalar}
\end{equation}
with the effective inverse metric
\begin{equation}
    g_{\rm eff}^{\mu\nu}=\Xi\,g^{\mu\nu}.
\label{eq:geff_scalar}
\end{equation}
The modification induced by the non-Riemannian geometry therefore corresponds to a purely conformal rescaling of the metric. Since Maxwell theory in four dimensions is conformally invariant, photon trajectories coincide with the usual null geodesics of the background metric.

Using \labelcref{eq:WaveCovector}, the dispersion relation \labelcref{eq:NullDispersion} implies
\begin{equation}
    k^\nu\mathring{\nabla}_\nu k^\mu=0,
\label{eq:ray_scalar}
\end{equation}
so that $k^\mu$ is tangent to affinely parametrized null geodesics of $g_{\mu\nu}$. If one instead introduces the conformally rescaled vector $k_{\rm eff}^\mu\equiv g_{\rm eff}^{\mu\nu}k_\nu=\Xi k^\mu$, then
\begin{equation}
    k_{\rm eff}^\nu\mathring{\nabla}_\nu k_{\rm eff}^\mu
    = \left(k_{\rm eff}^\alpha\mathring{\nabla}_\alpha\ln\Xi\right)
    k_{\rm eff}^\mu ,
\label{eq:autoparallel_scalar}
\end{equation}
which corresponds indeed to a non-affinely parametrized geodesic equation. The extra term is parallel to $k_{\rm eff}^\mu$ and therefore does not deflect the ray; it only reflects the reparametrization induced by the conformal factor.

\subsubsection*{GO limit: $\mathcal{O}(\omega)$}

The effects of torsion and non-metricity arise at the next order in the GO expansion. At $\mathcal{O}(\omega)$ one obtains
\begin{equation}
    2\,\Xi\,k^\alpha\mathring{\nabla}_\alpha a_\mu
    +\Xi\,\theta\,a_\mu
    +(a_\mu k^\alpha-a^\alpha k_\mu)\,\mathring{\nabla}_\alpha\Xi=0,
\label{eq:GO_Xi_omega1}
\end{equation}
with $\theta$ defined in \labelcref{eq:ThetaDefinition}. This equation governs the transport of the electromagnetic amplitude along the null rays.

To extract the evolution of the radiation intensity along the ray bundle, we decompose the amplitude as
\begin{equation}
    a_\mu=\mathcal{A}\,e_\mu,
\end{equation}
where $\mathcal{A}$ is a real scalar amplitude and $e_\mu$ is a unit polarization vector. Since the scalar model preserves the standard $U(1)$ gauge symmetry, one may consistently impose the Lorenz gauge condition $\mathring{\nabla}^\mu A_\mu=0$, which at leading GO order implies \cite{Poisson_Will_2014,Dolan:2017zgu,Maggiore:2007ulw} the transversality condition
\begin{equation}
    k^\mu e_\mu=0.
    \label{eq:GO_TransversalityCondition}
\end{equation}
Projecting \labelcref{eq:GO_Xi_omega1} along $e^\mu$, and using $e^\mu e_\mu=1$ together with \labelcref{eq:GO_TransversalityCondition}, yields
\begin{equation}
    2\,k^\alpha\mathring{\nabla}_\alpha\ln\mathcal{A}
    +\theta
    +k^\alpha\mathring{\nabla}_\alpha\ln\Xi
    =0 .
\label{eq:GO_Xi_amplitude}
\end{equation}
Defining the scalar intensity as
\begin{equation}
    \mathcal{I}\coloneqq\Xi\,\mathcal{A}^2,
\label{eq:intensity_definition_scalar}
\end{equation}
the transport equation becomes
\begin{equation}
    k^\alpha\mathring{\nabla}_\alpha\mathcal{I}
    +\theta\,\mathcal{I}=0 ,
\label{eq:Xi_intensity_transport}
\end{equation}
which has the standard GO form. In particular, writing
$\theta=\frac{1}{\Sigma}(d\Sigma/d\lambda)$ with $k^\alpha=dx^\alpha/d\lambda$ for the expansion of a ray bundle with cross section $\Sigma$, \labelcref{eq:Xi_intensity_transport} implies
\begin{equation}
    \mathcal{I}(\lambda)\,\Sigma(\lambda)=\text{const}.
\label{eq:intensity_conservation_scalar}
\end{equation}
Thus the non-Riemannian effects encoded in $\Xi$ do not violate the conservation of the dressed GO flux, but rather modify the quantity that is transported along the beam. The conserved intensity is $\mathcal{I}$ in \labelcref{eq:intensity_definition_scalar} instead of the bare squared amplitude, and as a consequence, an observer interpreting the signal in terms of the standard Maxwell normalization may infer an apparent flux renormalization, even though the canonically dressed intensity satisfies the usual conservation law. \revision{A similar modification of photon-number conservation, which leads to a violation of the distance-duality relation, was found for a scalar-field-dependent prefactor of the Maxwell term \cite{Hees:2014lfa} and for a non-minimal coupling to the non-metricity scalar of symmetric teleparallel gravity \cite{DAgostino:2026uct}. Here, $\Xi$ is instead built from the parity-even, projectively invariant quadratic scalars of $T_{\mathrm f}$ and $Q_{\mathrm f}$ in \labelcref{eq:Xi_general}.}

In summary, the scalar model is characterized by the fact that torsion and non-metricity enter only through an overall scalar dressing of the Maxwell sector. As a consequence, it does not deform the leading-order light cone and therefore preserves the standard null propagation of the wavefronts, while it modifies the relation between field amplitude and effective intensity. This makes the model particularly useful for isolating amplitude-transport effects without introducing, at leading order, polarization-dependent propagation or birefringence.

\subsection{Constitutive-tensor model}
\label{subsec:GO_constitutive_model}

We now turn to the constitutive-tensor model \labelcref{eq:Lchi_general}. In this case, the effect of the metric--affine geometry is no longer encoded in a simple scalar dressing of the Maxwell term, but in a rank--four tensor $\chi^{\mu\nu\alpha\beta}$ that modifies the tensorial structure of the kinetic sector. This opens the door to genuinely new effects, such as polarization-dependent propagation, birefringence, and mode mixing, which are absent in the purely scalar dressing model. 

It is convenient to introduce the excitation tensor
\begin{equation}
    H^{\mu\nu}\coloneqq\chi^{\mu\nu\alpha\beta}\mathring{F}_{\alpha\beta},
\label{eq:ExcitationTensor}
\end{equation}
so that the EoMs then take the compact form
\begin{equation}
    \mathring{\nabla}_\mu H^{\mu\nu}=0
    \quad\rightarrow\quad
    \chi^{\mu\nu\alpha\beta}\mathring{\nabla}_\mu\mathring{F}_{\alpha\beta} +(\mathring{\nabla}_\mu\chi^{\mu\nu\alpha\beta})\,\mathring{F}_{\alpha\beta}=0.
\label{eq:EOM_chi_general}
\end{equation}
In this language, the non-Riemannian geometry affects the dynamics both through the constitutive relation $H\sim\chi\cdot \mathring F$ and through the derivative term $\mathring{\nabla}\chi$ \cite{Novello:1999pg,Queiruga:2019ike}.
The latter term can be seen as an effective, not necessarily conserved current $J^\nu_{\rm eff}\sim (\mathring{\nabla}_\mu\chi^{\mu\nu\alpha\beta})\,\mathring{F}_{\alpha\beta}$ describing the exchange of energy and momentum between photons and the underlying geometry, causing an amplification or damping of wave amplitudes. The former term, instead, may lead to the transformation of photons into torsion/non-metricity and vice versa following a mechanism analogous to the Gertsenshtein effect \cite{Palessandro:2023tee}, by which graviton-photon oscillations are possible in regions of strong magnetic fields.

\subsubsection*{GO limit: $\mathcal{O}(\omega^2)$}

Inserting the GO Ansatz \labelcref{eq:GO_Ansatz} into \labelcref{eq:EOM_chi_general} and collecting the leading $\mathcal{O}(\omega^2)$ terms yields 
\begin{equation}
    k_\mu\chi^{\mu\nu\alpha\beta}(k_\alpha a_\beta-k_\beta a_\alpha)=0.
\label{eq:GO_chi_omega2}
\end{equation}

Defining the leading-order field strength and the corresponding excitation as
\begin{equation}
    f_{\alpha\beta}\coloneqq k_\alpha a_\beta-k_\beta a_\alpha, \qquad
    H^{\mu\nu}_{(0)}=\chi^{\mu\nu\alpha\beta}f_{\alpha\beta},
\label{eq:fandH0}
\end{equation}
\labelcref{eq:GO_chi_omega2} can be written simply as
\begin{equation}
    k_\mu H^{\mu\nu}_{(0)}=0.
\label{eq:GO_chi_kH}
\end{equation}

Equivalently, it can be interpreted as a polarization eigenvalue problem
\begin{equation}
    a_\alpha\,P\ud{\alpha}{\mu}(k,T,Q)=0,
    \label{eq:ConstitutiveLinearGeneral}
\end{equation}
whose reduced determinant, after removing the gauge null direction, defines the modified physical dispersion relation.
In general, the tensorial structure of $\chi^{\mu\nu\alpha\beta}$ leads to polarization-dependent propagation and therefore to birefringence \cite{Itin:2003hr,Novello:1999pg}.

\revision{For a general constitutive tensor, the characteristic system \labelcref{eq:GO_chi_omega2} has been studied in premetric electrodynamics \cite{Itin:2009aa}, where $\chi^{\mu\nu\alpha\beta}$ is taken as given and its derivatives are neglected. The resulting dispersion relation is quartic in $k_\mu$. Here, instead, $\chi^{\mu\nu\alpha\beta}$ is constructed from the projectively invariant parts of torsion and non-metricity, guided by $U(1)$ gauge and projective invariance, and its derivatives enter the amplitude transport at next-to-leading order in the GO expansion. Since our constitutive tensor is a small deformation of the Maxwell one, the quartic dispersion relation splits into two sheets close to the light cone, one for each polarization. To connect the general result with our models, we write \labelcref{eq:GO_chi_omega2} as
\begin{equation}
    M^{\nu\beta}a_\beta=0 , \qquad M^{\nu\beta}\coloneqq k_\mu k_\alpha\big(\chi^{\mu\nu\alpha\beta}-\chi^{\mu\nu\beta\alpha}\big) .
\label{eq:M_Itin}
\end{equation}
For amplitudes transverse to $k$, $k^\mu a_\mu=0$, this system coincides with \labelcref{eq:ConstitutiveLinearGeneral}, since $a_\alpha P\ud{\alpha}{\mu}=M\du{\mu}{\beta}a_\beta$. One has $k_\nu M^{\nu\beta}=M^{\nu\beta}k_\beta=0$ identically, so that $\det M=0$\revision{, as a consequence of gauge invariance,} and the adjugate matrix takes the form \cite{Itin:2009aa}
\begin{equation}
    \mathrm{adj}(M)_{\nu\beta}=\lambda(k)\,k_\nu k_\beta , \qquad \lambda(k)=\mathcal{G}^{\alpha\beta\gamma\delta}(\chi)\,k_\alpha k_\beta k_\gamma k_\delta ,
\label{eq:adjM_Itin}
\end{equation}
where $\mathcal{G}$ is cubic in $\chi^{\mu\nu\alpha\beta}$. The function $\lambda$ is\revision{, up to a factor that depends on how the gauge direction is removed,} the reduced determinant mentioned above, and $\lambda(k)=0$ defines the characteristic surfaces along which light propagates. For Maxwell theory, $\lambda\propto(k^2)^2$, i.e., the doubly degenerate light cone \labelcref{eq:NullDispersion}.}

\revision{Two properties of our models follow from \revision{\labelcref{eq:adjM_Itin}}. First, $\lambda$ is cubic and homogeneous in $\chi$, so an overall factor $\Xi$ only rescales it, $\lambda\to\Xi^3\lambda$, and drops out of $\lambda(k)=0$\revision{; for the scalar model, $\lambda\propto\Xi^3(k^2)^2$}. This is why the scalar model leaves the light cone unchanged, as found in \labelcref{eq:GO_Xi_omega2,eq:keff_scalar,eq:geff_scalar}. Second, the totally antisymmetric (axion) part $\chi^{[\mu\nu\alpha\beta]}$ drops out of $M^{\nu\beta}$ \cite{Itin:2009aa}. In \labelcref{eq:chi_general} this part arises only from the torsion term, in line with \cite{Itin:2003hr}. It therefore does not deform the light cone, although its derivatives still enter \labelcref{eq:GO_chi_omega1}, and the deformation comes from the remaining parts of the torsion and non-metricity terms.}

\revision{As noted above, $M^{\nu\beta}$ annihilates $k_\mu$, so that the direction $a_\mu\propto k_\mu$ is pure gauge and the physical polarizations are the amplitudes satisfying \labelcref{eq:GO_TransversalityCondition} modulo $k_\mu$ \cite{Poisson_Will_2014,Dolan:2017zgu,Maggiore:2007ulw}. For $k$ null with respect to $g$, this space is two-dimensional and spacelike, and $P$ is symmetric on it, since $\chi$ is symmetric under the exchange of its two index pairs. \revision{Explicitly, choosing an auxiliary null vector $l^\mu$ with $k_\mu l^\mu\neq0$, the projector onto this space is $h\ud{\mu}{\nu}=\delta^\mu_\nu-(k^\mu l_\nu+l^\mu k_\nu)/(k_\alpha l^\alpha)$. Different choices of $l^\mu$ change $h\ud{\mu}{\nu}$ only by terms proportional to $k$, so the results below do not depend on this choice.} To first order in $\bar c_T$ and $\bar c_Q$, the two sheets of the dispersion relation are obtained by restricting \labelcref{eq:ConstitutiveLinearGeneral} to this space, as in degenerate perturbation theory around the Maxwell light cone: the perturbation is evaluated on the Maxwell cone, while $k^2$ is kept to locate the shifted one.} For the explicit Ansatz \labelcref{eq:chi_general}, \revision{we therefore decompose the restricted operator into its trace and traceless parts,}
\begin{equation}
    P\ud{\alpha}{\mu}(k,T,Q) = \delta^\alpha_\mu\,\mathcal S(k,T,Q) + B\ud{\alpha}{\mu}(k,T,Q)\revision{,}
\label{eq:P_decomposition_explicit}
\end{equation}
\revision{where $\mathcal S$ is half the trace of $P$ on the space of physical polarizations and $B\ud{\alpha}{\mu}$ is traceless on it. The two polarization eigenvalues of $P$ are then $\mathcal S\pm b$, where $\pm b$ are the (real) eigenvalues of $B$\revision{, so that, to first order in the couplings, the two physical characteristic sheets are $\mathcal S\pm b=0$, and not simply $\mathcal S=0$; at this order, $\lambda\propto(\mathcal S+b)(\mathcal S-b)$}. Hence, $\mathcal S$ determines the polarization-averaged light cone, while $B$ controls the polarization dependence of the propagation and, for $b\neq0$, birefringence \cite{Itin:2003hr,Novello:1999pg}.}

\revision{To first order, \revision{writing $M^{\nu\beta}=k^2g^{\nu\beta}-k^\nu k^\beta+\delta M^{\nu\beta}$, one has $\mathcal S=k^2+\tfrac12\,h_{\nu\beta}\,\delta M^{\nu\beta}$, with $\delta M^{\nu\beta}$ evaluated on the Maxwell cone. Since $\delta M^{\nu\beta}k_\beta=0$,} this trace can be computed directly with the full metric, $\mathcal S=\tfrac12\,g_{\nu\beta}M^{\nu\beta}-\tfrac12\,k^2$. For \labelcref{eq:chi_general} this gives}
\begin{align}
    \color{revcolor}\mathcal S(k,T,Q) &\color{revcolor}= k^2 + \bar{c}_T\,k_\mu k_\nu\,T_{\mathrm f}^{\;\delta\mu\sigma}\,T\du{\mathrm{f},\,\delta}{\nu}{}_{\sigma} \nonumber\\[1ex]
    &\color{revcolor}\quad - \frac{1}{2}\,\bar{c}_Q\,k_\mu k_\nu\,Q_{\mathrm f}^{\;\delta\mu\sigma}\,Q\du{\mathrm{f},\,\delta}{\nu}{}_{\sigma} ,
\label{eq:Sterm_TQ_def}
\end{align}
\revision{which, like $B$, is built from $T_{\mathrm f}$ and $Q_{\mathrm f}$ only and is therefore projectively invariant. The polarization-averaged light cone is thus described by} an effective inverse metric in the spirit of \labelcref{eq:keff_scalar}, with
\begin{equation}
    \color{revcolor}g_{\rm eff}^{\mu\nu} \color{revcolor}= g^{\mu\nu} + \bar{c}_T\,T_{\mathrm f}^{\;\delta\mu\sigma}\,T\du{\mathrm{f},\,\delta}{\nu}{}_{\sigma}
    \color{revcolor} - \frac{1}{2}\,\bar{c}_Q\,Q_{\mathrm f}^{\;\delta\mu\sigma}\,Q\du{\mathrm{f},\,\delta}{\nu}{}_{\sigma} .
\label{eq:geff_TQ_def}
\end{equation}
\revision{Beyond first order in the couplings, the exact characteristic surface is the quartic $\lambda(k)=0$ of \cite{Itin:2009aa}, from which the two sheets follow exactly. The corrections to the first-order result are of fourth order in $T_{\mathrm f}$ and $Q_{\mathrm f}$, beyond the order of the Ansatz \labelcref{eq:chi_general_symmetric}, and are therefore not considered here.}

In contrast with the scalar-coupled model, \revision{this effective metric} is not simply conformal to $g^{\mu\nu}$, since torsion and non-metricity also generate anisotropic corrections. Therefore, already the \revision{trace part $\mathcal S$} of the constitutive model can modify the characteristic surfaces themselves, rather than only the parametrization of the rays.

The tensorial piece $B$ is linear in $(\bar{c}_T,\bar{c}_Q)$ and quadratic in \revision{$(T_{\mathrm f},Q_{\mathrm f})$} and encodes the polarization-dependent corrections to the leading GO propagation. Its explicit form is rather lengthy and will not be needed in full generality below, but a detailed evaluation of this tensorial sector for specific metric--affine backgrounds will be addressed in future phenomenological partner papers.

Thus, in general, the constitutive coupling leads to modified light cones, birefringence and polarization-dependent propagation, consistently with the effective-medium interpretation of the model.

\subsubsection*{GO limit: $\mathcal{O}(\omega)$}

At linear order in $\omega$ one finds
\begin{align}
    k_\mu\chi^{\mu\nu\alpha\beta}
    (\mathring{\nabla}_\alpha a_\beta-\mathring{\nabla}_\beta a_\alpha) 
    + \mathring{\nabla}_\mu H^{\mu\nu}_{(0)}=0 .
\label{eq:GO_chi_omega1}
\end{align}
This is the analog of the amplitude transport equation \labelcref{eq:GO_Xi_omega1} obtained in the scalar model. Since the constitutive tensor acts as an anisotropic effective medium, its structure is richer and the evolution of the polarization generally involves mixing between the two transverse modes. The detailed form of this transport law therefore depends on the explicit structure of $\chi^{\mu\nu\alpha\beta}$ and hence on the torsion and non-metricity background.

Without specifying a particular background, a useful way to display the structure of the transport equation is to work locally in the eigenbasis of the constitutive map acting on two-forms. For a fixed wave covector $k_\mu$, we introduce the constitutive operator
\begin{equation}
    (\chi\cdot f)^{\mu\nu}
    \coloneqq
    \chi^{\mu\nu\alpha\beta}f_{\alpha\beta}.
\label{eq:ConstitutiveOperator}
\end{equation}
If one works locally in a polarization eigenbasis, the leading-order field strength may be taken, schematically, to satisfy
\begin{equation}
    (\chi\cdot f)^{\mu\nu}
    =\Xi_\chi\,f^{\mu\nu},
\label{eq:chi_eigenform}
\end{equation}
where $\Xi_\chi=\Xi_\chi(k,T,Q,e)$ is the corresponding eigenvalue. In terms of the excitation tensor this is simply
\begin{equation}
    H^{\mu\nu}_{(0)}=\Xi_\chi\,f^{\mu\nu}.
\label{eq:H0_eigenform}
\end{equation}
\revision{This form is, however, restrictive. Inserting \labelcref{eq:H0_eigenform} into \labelcref{eq:GO_chi_kH} and using $k^\mu a_\mu=0$ gives $\Xi_\chi\,k^2a^\nu=0$, so that a nontrivial physical mode of this type has $k^2=0$. The simplified \revision{intensity} law below therefore holds only for modes that propagate on the undeformed light cone, as in the scalar limit $\chi^{\mu\nu\alpha\beta}=\Xi\,g^{\mu[\alpha}g^{\beta]\nu}$. It does not apply to generic birefringent rays on the deformed sheets $\mathcal S\pm b=0$.}
\revision{Moreover, \labelcref{eq:H0_eigenform} fixes $\chi^{\mu\nu\alpha\beta}$ on $f_{\alpha\beta}$ but not on the amplitude derivatives in \labelcref{eq:GO_chi_omega1}, so it does not reduce \labelcref{eq:GO_chi_omega1} to the form of \labelcref{eq:GO_Xi_omega1}. Instead, contracting \labelcref{eq:GO_chi_omega1} with $a_\nu$ and using \labelcref{eq:chi_symmetries} gives, without further assumptions,}
\begin{equation}
    \color{revcolor}\mathring{\nabla}_\mu\mathcal{J}^\mu=0 , \qquad \mathcal{J}^\mu\coloneqq a_\nu H^{\mu\nu}_{(0)} .
\label{eq:GO_chi_transport_vec}
\end{equation}
\revision{For an eigenmode of the form \labelcref{eq:H0_eigenform} that can be followed independently, \revision{a real amplitude} $a_\mu=\mathcal{A}\,e_\mu$ gives $\mathcal{J}^\mu=\Xi_\chi\,\mathcal{A}^2k^\mu$.} One may then define the corresponding mode intensity in an analogous way to \labelcref{eq:intensity_definition_scalar} as
\begin{equation}
    \mathcal{I}\coloneqq\Xi_\chi\,\mathcal{A}^2 .
\label{eq:intensity_definition_constitutive}
\end{equation}
Thus, \revision{for modes of this type,} the product of the intensity and the beam cross section is conserved \revision{along the ray}, while the dependence on torsion and non-metricity is encoded in the effective constitutive eigenvalue $\Xi_\chi$.

In a generic polarization basis, however, $H^{\mu\nu}_{(0)}$ is not proportional to $f^{\mu\nu}$ and the linear GO equation does not reduce to a single scalar transport law. Instead, the polarization amplitudes form a vector in the two-dimensional space of transverse modes, and their evolution along the ray is governed schematically by
\begin{equation}
    k^\alpha \mathring{\nabla}_\alpha a^{(i)}
    = \mathcal{M}\ud{(i)}{(j)}(k,T,Q)\,a^{(j)},
\label{eq:polarization_mixing_transport}
\end{equation}
where $\mathcal{M}\ud{(i)}{(j)}$ encodes the polarization mixing induced by the anisotropic structure of $\chi^{\mu\nu\alpha\beta}$. The simpler scalar-like \revision{intensity law \labelcref{eq:intensity_definition_constitutive}} is recovered only after choosing the local eigenbasis of the constitutive map, where the two polarization modes are diagonalized and can be followed independently\revision{, and only for the modes on the undeformed light cone identified after \labelcref{eq:H0_eigenform}. The polarization-resolved transport along the deformed sheets\revision{, where rays need not follow $k^\alpha$,} is left for future work}.

The phenomenological implications of the constitutive-tensor model are thus potentially much richer than in the standard framework of GR with Maxwell electrodynamics because the spacetime would no longer be insensitive to polarization, becoming a complex medium. In particular, the optical appearance of black holes could be dramatically affected by birefringence effects. Different propagation speeds for photon polarizations (induced by torsion and non-metricity) would lead to the superposition of two images, leaving a clear imprint on the inner shadow. In GR, the polarization plane of photons rotates along geodesic trajectories, but the polarization degree is conserved. In our case, depolarization may occur due to energy leakage from the electromagnetic fields to torsion and non-metricity, inducing a clear contrast between the light coming from the back side of the accretion disk and that emitted from the front, which crosses weaker torsion and non-metricity gradient regions. If the constitutive tensor breaks time-reversal invariance, a contrast between co-rotating and counter-rotating photons could emerge on top of the usual frame-dragging effect, making the shadow more asymmetric than expected in the Kerr scenario. One cannot exclude the possibility of geometric light modulation due to the transmutation of photons into torsion/non-metricity and vice versa \cite{Palessandro:2023tee}.

The cosmological traces of the new couplings presented here are also of the utmost importance. \revision{Potential imprints include 1) in gravitational lensing, the dislocation of Einstein rings by massive structures or quasar time delays due to birefringence, 2) characteristic polarization patterns in the cosmic microwave background (CMB) induced by a pseudoscalar (axion-like) coupling in $\chi^{\mu\nu\alpha\beta}$, and 3) \revision{apparent} photon-number loss due to $\mathring{\nabla}_\mu\chi^{\mu\nu\alpha\beta}$ couplings to other sources, possibly affecting the interpretation of luminosity distances, which assume photon conservation.} 

\section{Conclusions}
\label{sec:Conclusions}

In this work we have studied the propagation of electromagnetic radiation in metric--affine geometries, where the affine connection is treated as an independent object and torsion and non-metricity may be present. In such geometries, the coupling between electromagnetism and spacetime is no longer unique, and different prescriptions for defining the field strength or for introducing covariant derivatives can lead to inequivalent EoMs. Rather than choosing one such prescription from the outset, we have adopted a symmetry-guided approach based on two requirements: preservation of the standard $U(1)$ gauge symmetry of Maxwell theory and invariance under projective transformations of the affine connection.

These two principles naturally restrict the admissible electromagnetic couplings, in the sense that gauge invariance selects the standard Levi--Civita Faraday tensor, while projective invariance requires the non-Riemannian sector to enter through projectively invariant combinations of torsion and non-metricity. Within this framework, we have introduced two representative classes of models: the first one is a scalar-coupled model, in which the Maxwell kinetic term is multiplied by a scalar prefactor $\Xi$ built from projectively invariant geometric quantities; the second one is a constitutive-tensor model, in which the kinetic term is modified by a rank--four tensor $\chi^{\mu\nu\alpha\beta}$, allowing the non-Riemannian geometry to act as an effective medium for the electromagnetic field.

We have then analyzed the GO limit of both models. In the scalar-coupled case, we have shown that the leading-order light cone is not deformed, and the effect of the non-Riemannian geometry can be interpreted as a conformal rescaling of the effective inverse metric, so that the wavefronts propagate along the usual null directions of the background metric. The scalar prefactor nevertheless enters at the next order in the GO expansion, modifying the relation between the field amplitude and the effective intensity. In particular, the conserved quantity along a ray bundle is not simply the squared amplitude, but the dressed intensity $\mathcal{I}=\Xi\mathcal{A}^2$. This type of effect may leave phenomenological signatures in observables sensitive to flux normalization, such as luminosity distances, shadow intensity profiles, or any setting in which photon-number conservation is effectively assumed.

For the constitutive-tensor model, torsion and non-metricity do not merely rescale the Maxwell sector, but also modify the tensorial structure of the electromagnetic kinetic term, making it qualitatively richer from the phenomenological perspective. At leading order in GO, the propagation equation becomes a polarization-dependent eigenvalue problem; \revision{its trace part, which gives the polarization-averaged light cone,} can be interpreted in terms of an effective inverse metric, whose anisotropic corrections may deform the light cone, while the remaining tensorial part controls polarization-dependent propagation, birefringence, and mode mixing. At the next order, the derivative term $\mathring{\nabla}\chi$ contributes to the transport of the amplitude and can be understood as an effective exchange between the electromagnetic field and the non-Riemannian background. This opens the possibility of phenomenological signatures in polarization observables, birefringent lensing, black-hole imaging, and other situations in which different polarization modes probe different effective optical geometries.

Altogether, our results show that torsion and non-metricity can affect light propagation in several distinct ways: by modifying the effective light cone, by changing the relation between amplitude and intensity, by inducing polarization-dependent propagation, and by generating mixing between polarization modes. These effects are absent in standard Maxwell theory on a Riemannian background and thus provide a direct route to connect metric--affine geometry with electromagnetic observables. Their quantitative impact is likely to be strongly model-dependent, but their implications for the understanding of gravitational interactions require an exhaustive analysis and the definition of strategies to isolate specific effects in different scenarios. We plan to develop such a research program in future works.

The present work therefore sets the formal basis for a systematic connection between metric--affine geometry and electromagnetic observations. By identifying the propagation channels through which torsion and non-metricity can affect light, it provides a concrete route from non-Riemannian geometry to phenomenology. The next papers in this series will apply this framework to specific metric--affine backgrounds and determine how their geometric structure is imprinted on observables such as black-hole images, birefringent lensing, polarization-dependent propagation, and \revision{apparent} departures from photon-number conservation.


\begin{acknowledgments}
This work is supported by the Spanish Grant PID2023-149560NB-C21 and the Severo Ochoa Excellence Grant CEX2023-001292-S, funded by MICIU/AEI/10.13039/501100011033 (“ERDF A way of making Europe”, “PGC Generación de Conocimiento”) and FEDER, UE. Support from CosmoVerse CA21136 COST action, European Cooperation in Science and Technology is also acknowledged. This work has also been supported by the European Horizon Europe staff exchange (SE) programme HORIZON-MSCA-2021-SE-01 Grant No. NewFunFiCO-101086251.
\end{acknowledgments}


\bibliographystyle{apsrev4-2}
\bibliography{biblio}   

\end{document}